\documentclass[letterpaper, 10 pt, conference]{ieeeconf}  

\IEEEoverridecommandlockouts                              
\usepackage{graphics} 
\usepackage{epsfig} 
\usepackage{mathptmx} 
\usepackage{times} 
\usepackage{amsmath} 
\usepackage{amssymb}  
\usepackage{color}
\usepackage{xcolor}

\title{{\sc DeepTurbo} : Deep Turbo Decoder 
}

\author{ \parbox{2 in}{\centering Yihan Jiang\\
	ECE Department\\
         University of Washington\\
	Seattle, United States \\
	{\tt\small yij021@uw.edu}}
	 \parbox{2 in}{\centering Hyeji Kim\\
	Samsung AI Center Cambridge\\
         Cambridge, United Kingdom\\
	{\tt\small hkim1505@gmail.com}}
	 \parbox{2 in}{\centering Himanshu Asnani\\
	ECE Department\\
         University of Washington\\
         Seattle, United States \\
	{\tt\small asnani@uw.edu}}\\
	\\
	 \parbox{2 in}{\centering Sreeram Kannan\\
	ECE Department\\
         University of Washington\\
         Seattle, United States \\
	{\tt\small ksreeram@ee.washington.edu}}
	 \parbox{2 in}{\centering Sewoong Oh\\
	Allen School of Computer Science $\&$ Engineering\\
         University of Washington\\
         Seattle, United States \\
	{\tt\small sewoong@cs.washington.edu}}
	 \parbox{2 in}{\centering Pramod Viswanath\\
	ECE Department\\
         University of Illinois at Urbana Champaign\\
         Illinois, United States\\
	{\tt\small pramodv@illinois.edu}}
}

\begin{document}

\maketitle
\thispagestyle{empty}
\pagestyle{empty}

\begin{abstract}



Present-day communication systems routinely use codes that approach the channel capacity when coupled with a computationally efficient decoder. However, the decoder is typically designed for the Gaussian noise channel, and is known to be sub-optimal for non-Gaussian noise distribution. Deep learning methods offer a new approach for designing decoders that can be  trained and tailored for arbitrary channel statistics. We focus on Turbo codes, and propose ({\sc DeepTurbo}), a novel deep learning based architecture for Turbo decoding. 

The standard Turbo decoder ({\sc Turbo}) iteratively applies the Bahl-Cocke-Jelinek-Raviv (BCJR) algorithm with an interleaver in the middle. A neural architecture for Turbo decoding, termed ({\sc NeuralBCJR}), was proposed recently. There, the key idea is to create a module that {\em imitates} the BCJR algorithm using supervised learning, and to use the interleaver architecture along with this module, which is then fine-tuned using end-to-end training. However, knowledge of the BCJR algorithm is required to design such an architecture, which also constrains the resulting learnt decoder. Here we remedy this requirement and propose a fully end-to-end trained neural decoder - \textit{Deep Turbo Decoder} ({\sc DeepTurbo}). With novel learnable decoder structure and training methodology, {\sc DeepTurbo} reveals superior performance under both AWGN and non-AWGN settings as compared to the other two decoders - {\sc Turbo}  and {\sc NeuralBCJR}. Furthermore, among all the three, {\sc DeepTurbo} exhibits the lowest error floor.

\end{abstract}


\section{Introduction}
\label{intro}

\subsection{Motivation}
Communication standards typically fix an encoder as a capacity-approaching code, while allowing different stakeholders to implement their own decoders~\cite{richardson2008modern}. Thus, designing decoding algorithms with properties such as \emph{high reliability}, \emph{robustness}, \emph{adaptivity} is of utmost interest to both industry and academia. 
Among several standard codes, turbo code is widely used in modern communication systems where the standard Turbo decoder ({\sc Turbo}) uses the iterative BCJR~\cite{bcjr} algorithm with an interleaving decoding procedure. Historically, this became the first channel coding scheme which achieves capacity-approaching performance under AWGN channels~\cite{berrou1993near}. Despite the near-optimal performance on the AWGN channel, {\sc Turbo} lacks both robustness and adaptivity on non-AWGN settings. Robustness refers to the upholding of the performance even at the face of unexpected noise.  
Since {\sc Turbo} is an iterative decoding algorithm with interleaving, 
a log-likelihood corrupted by bursty noise on one coded bit can propagate to other coded bits via iterative decoding, which ends up with severely degraded decoding performance~\cite{kim2018communication}. 
Adaptivity refers to the capability to adapt exhibiting competing performance over a number of different channels, when the channel statistics are known to the decoder.  
Traditional methods improve adaptivity via whitening and thresholding the received signal with heuristics, so as to make the AWGN-pre-designed decoder work well~\cite{richardson2008modern}. Heuristic algorithms show different levels of performance, but there is no guarantee on the performance under unexpected settings~\cite{li2013ofdma}. Moreover, Turbo's error floor refers to the flattened decoding performance on the high signal-to-noise ratio (SNR) region. As a result, Turbo code is not viable for applications which require high reliability such as secure communications and authentications~\cite{richardson2003ldpcerr}~\cite{perez1996err}. The existence of the Turbo error floor relies on the low-weight codeword distributions~\cite{perez1996err}. Multiple error floor lowering techniques have been proposed since the inception of Turbo code. Annealing methods transform the posterior of each stage to avoid transient chaos ~\cite{kocarev2002turbo}. Designing a better interleaver to avoid low-weight codewords is proposed in~\cite{bohorquez2015interleaver}. Concatenating another outer code such as the Bose-Chaudhuri-Hocquenghem (BCH) code is also widely used~\cite{andersen1996outercode}. Applying additional check code such as cyclic redundancy check (CRC) code for post-processing is proposed in~\cite{leanderson2005}~\cite{o2007imp}~\cite{tonnellier2016l}. Error floor lowering methods show different levels of success with handcrafted heuristics. In summary, the traditional design of {\sc Turbo} with heuristics suffer to deliver on the desired features such as high reliability, robustness, adaptivity, and lower error floor. 

\subsection{Prior Art}
Using deep learning based methods for channel coding has received sufficient attention since its inception~\cite{o2016learning}~\cite{o2017introduction}. In particular, deep learning based decoders for canonical channel codes exhibit competing performance. BCH and High-Density Parity-Check (HDPC) codes can be decoded near optimally via a learnable Belief Propagation (BP) decoder~\cite{nachmani2016learning}~\cite{nachmani2018deep}. Polar codes can be also decoded by neural BP~\cite{gruber2017deep}~\cite{cammerer2017scaling}. Neural decoding for convolutional code and turbo code was introduced in ~\cite{kim2018communication}, where in the case of turbo code, the introduced iterative neural BCJR decoder ({\sc NeuralBCJR}) starts by imitating BCJR algorithm with RNN. It then equips the decoding with adaptivity by replacing BCJR algorithm with a pre-trained RNN followed by a further end-to-end fine turning of the parameters for non-AWGN settings. {\sc NeuralBCJR} shows high reliability, with matching {\sc Turbo} performance under different block lengths and signal-to-noise-ratio (SNR) on AWGN channel. 
On non-AWGN channels, {\sc NeuralBCJR} is robust to unexpected noise, and can further be trained to adapt to other channel settings. 
{\sc NeuralBCJR} imitates {\sc Turbo} with RNN, thereby also inheriting its disadvantages, viz. the error floor. Furthermore, BCJR pre-training is required for training {\sc NeuralBCJR}. A natural question to as here is : \textit{Can we train a neural Turbo decoder without BCJR knowledge to get even better performance?}

\subsection{Our Contribution}

In this paper, we answer the above question in affirmative and introduce Deep Turbo Decoder ({\sc DeepTurbo}), the first end-to-end learnt capacity-approaching neural Turbo decoder without BCJR knowledge. {\sc DeepTurbo} has high reliability with respect to bit error rate (BER) and block error rate (BLER), robustness and adaptivity on various channels, and lower error floor as compared to {\sc Turbo} and {\sc NeuralBCJR}. {\sc DeepTurbo} also achieves desired BER performance with reduced decoding iterations. The paper is organized as follows: 

Section \ref{turbofy} introduces the structure and training algorithm for {\sc DeepTurbo}, while Section \ref{perf} examines its performance on AWGN and non-AWGN channels. 
The paper concludes with discussion on the open issues and future directions in Section \ref{discussion}.

\section{{\sc DeepTurbo} Architecture}
\label{turbofy}

\subsection{Turbo Encoder}
Turbo encoder is composed of an interleaver and recursive systematic convolutional (RSC) encoders as shown in  Figure \ref{turbo}, where the interleaver ($\pi$) shuffles the input string with a given order; while the deinterleaver ($\pi^{-1}$) undos the interleaving in a known order. RSC code with generating function $(1, \frac{f_1(x)}{f_2(x)})$ servers as the encoding block for Turbo code. Two commonly used configurations of RSC are used in this paper:
\begin{itemize}
    \item code rate $R=1/3$, with $f_1(x) = 1 + x^2$ and $f_2(x) = 1+x+x^2$, which is denoted as turbo-757.
    \item code rate $R=1/3$, with $f_1(x) = 1 + x^2 +x^3$ and $f_2(x) = 1+x+x^3$, which is standard Turbo code used in LTE system, denoted as turbo-LTE.
\end{itemize}

\noindent\textbf{Notation.} A rate 1/3 turbo encoder generates three coded bits ($x_1, x_2, x_3$) per each message bit. As illustrated in Figure~\ref{turbo} (up), among the three coded bits, first output $x_1$ is the systematic bit, and $x_2$ is a coded bit generated through an RSC, and $x_3$ is a coded bit generated through an RSC for the interleaved bit stream. We let $(y_1, y_2, y_3)$ denote the noise corrupted versions of $(x_1, x_2, x_3)$ that the decoder receives.

\subsection{Turbo Decoders}
\begin{figure}[!ht] 
\centering
\includegraphics[width=0.17\textwidth]{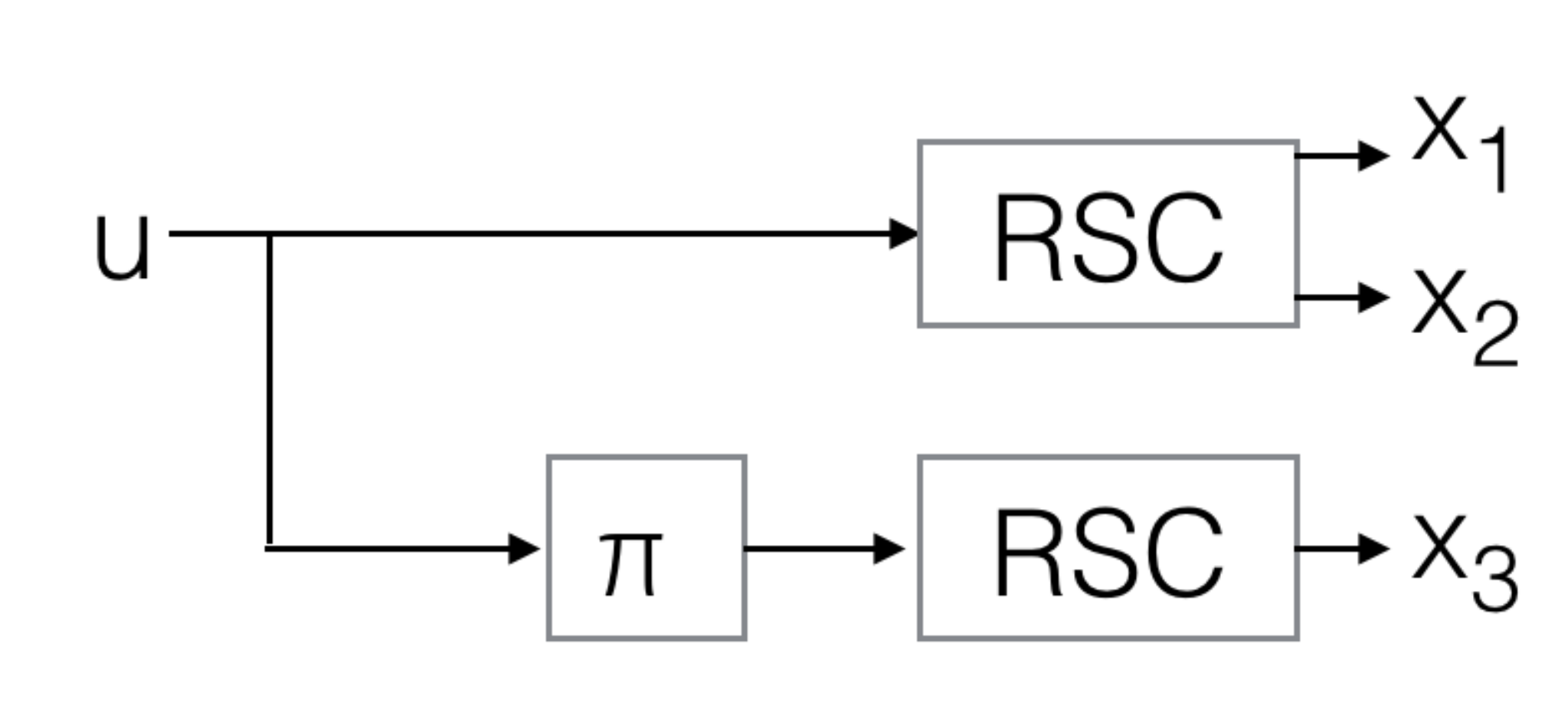}\ \ \ 
\includegraphics[width=0.48\textwidth]{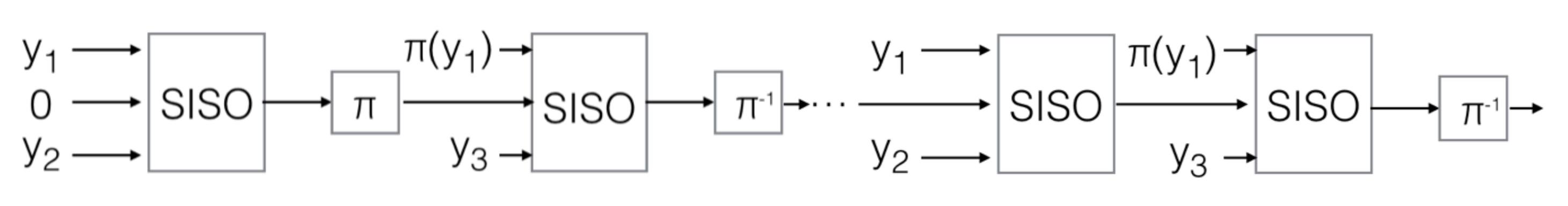}\ \ \ 
\caption{Turbo Encoder (up) and Decoder (down) }\label{turbo}
\vspace{-0.5em}
\end{figure}

Turbo decoders are designed under the `Turbo Principle'~\cite{hagenauer1997turbo}, which is iteratively refining posterior information by interleaved/de-interleaved received signals with soft-in soft-output (SISO) decoders. SISO decoder takes received signals and prior, and produces posterior as prior for later SISO blocks. The general Turbo decoder with rate 1/3 structure is shown in Figure \ref{turbo} (down). Each decoding iteration uses two SISO decoders to decode, the first stage takes de-interleaved posterior $p$ from last stage as prior, and received signal $y_1$ and $y_2$ as inputs; while the second stage takes interleaved posterior $\pi(q)$ as prior and received signal $\pi(y_1)$ and $y_3$. SISO outputs posterior $q$ to be fed to the next stage. At the end of iterative decoding procedure, decoding is done according to the estimated posterior. 

\subsection{Design of SISO Algorithm}
Different Turbo decoding algorithms differ in the design of the SISO algorithm. We compare the SISO design of the {\sc Turbo}, {\sc NeuralBCJR}, and our proposed {\sc DeepTurbo}:

\subsubsection{Standard Turbo Decoder ({\sc Turbo})} {\sc Turbo}~\cite{berrou1993near} uses the BCJR algorithm for the SISO algorithm with extrinsic information and noise weighted systematic bits, cf. Figure \ref{decs} (left). The extrinsic information takes the difference between the prior and the posterior to be fed to the next stage. {\sc Turbo} also requires estimating the channel noise variance to compensate noise. The posterior is compensated with extrinsic information and compensated systematic bits, before it is sent to next stage as prior. While {\sc Turbo} is designed to operate reliably under AWGN settings, it is sensitive to non-AWGN noises as in a non-AWGN setting. For example, a bursty noise corrupted bit leads to severely degraded performance (shown in Figure \ref{better_adapt}), since the iterative decoding scheme propagates corrupted posterior to other code bits via interleaving. Even under AWGN channel, {\sc Turbo} suffers from the error floor due to the existence of low weight codewords.

\subsubsection{Iterative Neural BCJR Decoder ({\sc NeuralBCJR})~\cite{kim2018communication}} {\sc NeuralBCJR} replaces the BCJR algorithm with Bidirectional Gated Recurrent Unit (Bi-GRU) (see appendix for more details). The SISO block of {\sc NeuralBCJR} is shown in Figure \ref{decs} (right), where it is Initialized by pre-trained Bi-GRU with BCJR input and output to imitate BCJR algorithm followed by an end-to-end training till convergence. All SISO blocks share the same model weights. {\sc NeuralBCJR} SISO block removes the link of compensating systematic bits and does not require estimating the channel noise to decode. {\sc NeuralBCJR} avoids producing inaccurate noise weighted systematic bits by implicitly estimating the channel, which improves the robustness against unexpected non-AWGN channels.  {\sc NeuralBCJR} shows matched performance on AWGN channel compared to {\sc Turbo} under AWGN channels, while it shows better robustness and adaptivity comparing to {\sc Turbo} with heuristics. However, the error floor of {\sc NeuralBCJR} is still high. In this design, there are two major caveats: (1) {\sc NeuralBCJR} requires to have BCJR knowledge to initialize the SISO block. Without BCJR-initialization, directly training {\sc NeuralBCJR} from scratch is not stable and (2) {\sc NeuralBCJR} simply replaces the BCJR block by weight-sharing Bi-GRUs, with the same input and output relationship, which limits its potential capacity of {\sc NeuralBCJR}.

\subsubsection{Deep Turbo Decoder ({\sc DeepTurbo})} 
\begin{figure}[!ht] 
\centering
\includegraphics[width=0.40\textwidth]{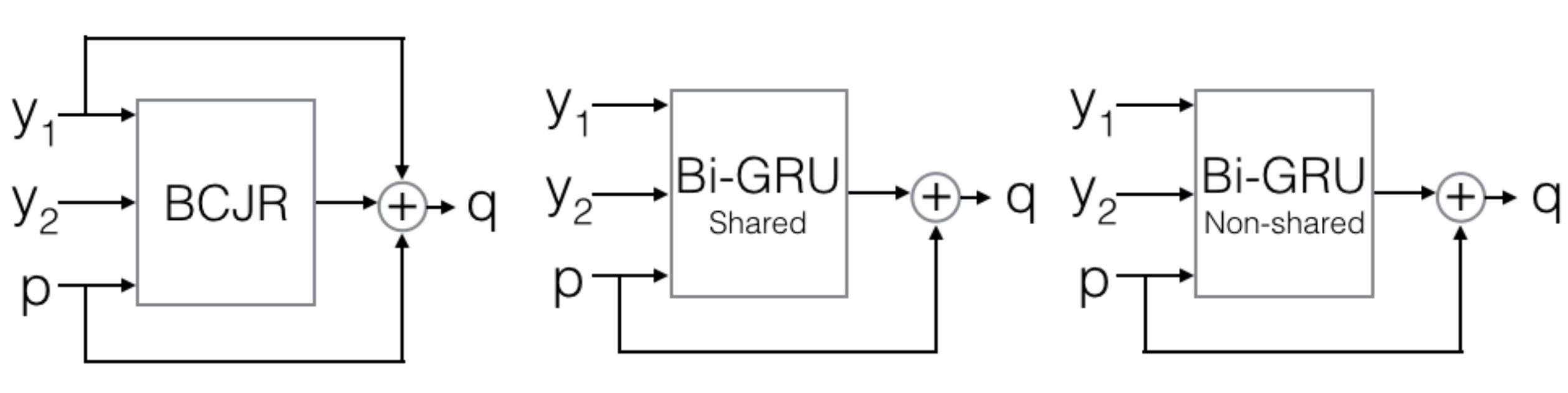}\ \ \ 
\caption{Different SISOs: {\sc Turbo} (left), {\sc NeuralBCJR} (middle) and {\sc DeepTurbo} (right)}\label{decs}
\vspace{-0.5em}
\end{figure}

To ameliorate the caveats of {\sc NeuralBCJR}, we propose {\sc DeepTurbo} where each SISO block (cf. Figure \ref{decs} (right)) still uses Bi-GRU as the building block, while keeping the extrinsic connection as a short-cut for gradient inspired by ResNet~\cite{he2016resnet}. 

Two major structural differences between {\sc NeuralBCJR} and {\sc DeepTurbo}:
\begin{itemize}
\item Non-shared weights. Unlike {\sc NeuralBCJR} uses the same Bi-GRU for all SISO blocks, {\sc DeepTurbo} doesn't share weight across different iterations, which allows each iteration to deal with posterior differently. Furthermore, non-shared weights improve the training stability.

\item More information passed to the next stage. Both {\sc Turbo} and {\sc NeuralBCJR} represent the posterior of each code bit by a single value log-likelihood (LLR). A single value for each code bit might not be sufficient to convey enough information. Inspired by resolving calibration issue by ensemble methods~\cite{Lakshminarayanan2017calibration}, for each code bit position, we take length $K$ bits instead of 1 bit. For example, for block length $L=100$, {\sc NeuralBCJR} posterior LLR has shape $(L, 1)$, while {\sc DeepTurbo} transmits a posterior of shape $(L, K)$ to next stage.

\end{itemize}

A significant advantage is that {\sc DeepTurbo} does not unitilize BCJR knowledge at all, which allows {\sc DeepTurbo} to learn a better decoding algorithm in a data-driven end-to-end approach. The hyper-parameters for {\sc DeepTurbo} decoder are shown in Figure~\ref{hyper}, further discussion is deferred to the appendix.

\begin{figure}[!ht] 
\centering
{\tiny
 \begin{tabular}{|c c|} 
 \hline
 {\sc DeepTurbo} SISO &  2-layer Bi-GRU with 100 units \\ [0.5ex] 
{\sc DeepTurbo-CNN} SISO & 5-layer 1d CNN with 100 filters, kernel size = 5, stride=1, padding=2 \\ 
 Learning rate & 0.001, decay by 10 when saturate \\
 Num epoch & 200\\
 Block length & 100/1000 \\
 Batch per epoch & 100\\
 Optimizer & Adam\\
 Loss  & BCE\\
Train SNR & -1.5dB\\
Batch size & 500\\ 
Posterior feature size K & 5\\
Decode iterations & 6\\
 [1ex] 
 \hline
\end{tabular}
\centering
\caption{{\sc DeepTurbo} hyperparameters}\label{hyper}
}
\end{figure}

\section{Deep Turbo Decoder performance}
\label{perf}
We compare the performance  of Deep Turbo Decoder ({\sc DeepTurbo}) with the baseline decoders, the standardTurbo decoder ({\sc Turbo}) and Iterative Neural BCJR Decoder ({\sc NeuralBCJR}), under both AWGN and non-AWGN channels.

\begin{figure}[!ht] 
\centering
\vspace{-0.5em}
\centerline{\includegraphics[width=0.49\textwidth]{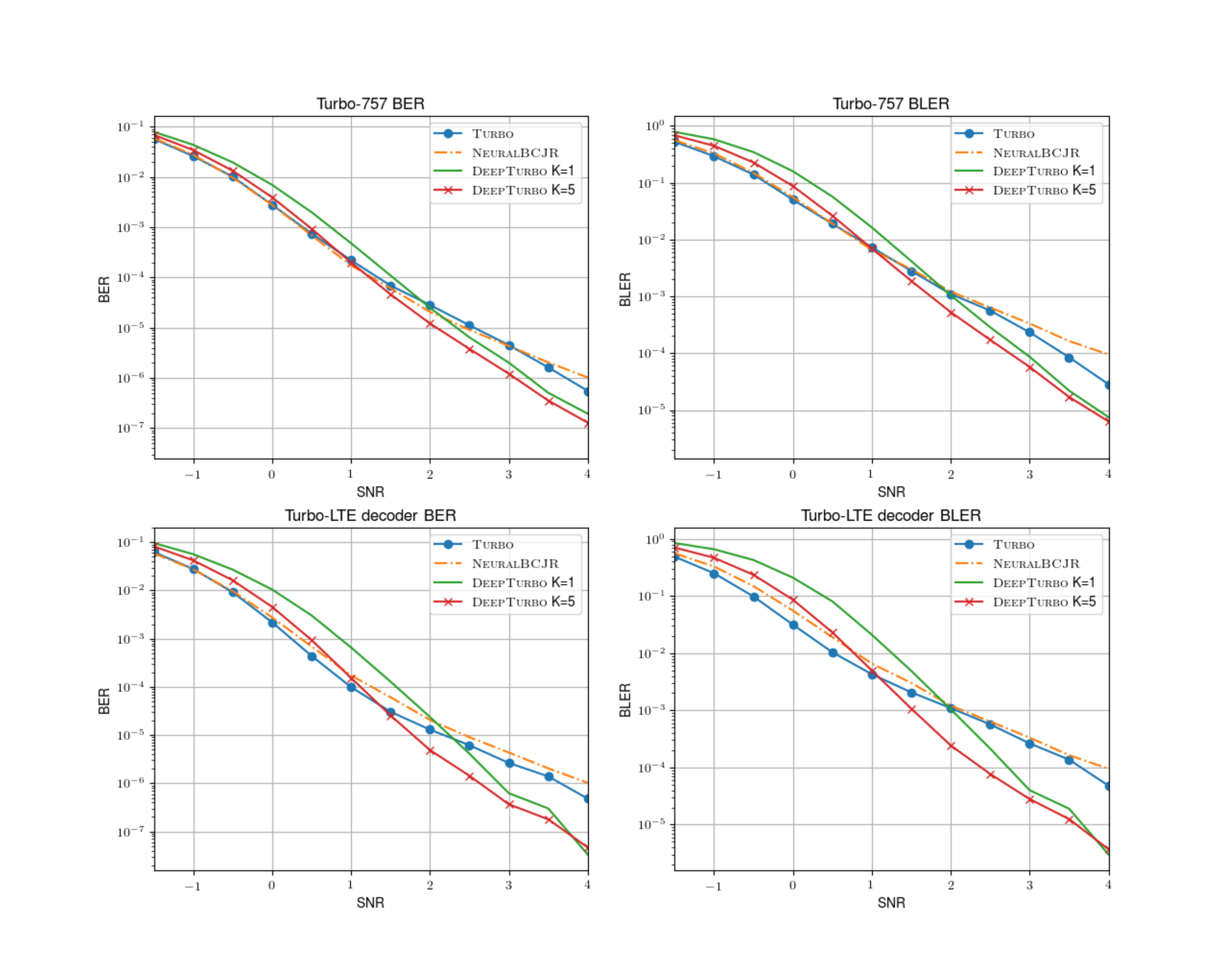}}\ \ \ 
\vspace{-1.5em}
\caption{The proposed Deep Turbo Decoder ({\sc DeepTurbo}) improves upon the standard Turbo decoder in the large SNR regime for Turbo-757 (up) and Turbo-LTE (down)}\label{bl100}
\vspace{-1.0em}
\end{figure}

\subsection{AWGN with a block length 100} 
In Figure~\ref{bl100}, we 
compare the decoder performances for Turbo codes with both turbo-757 and turbo-LTE, trained under block length 100, and decoding iteration 6. 
{\sc NeuralBCJR} matches the performance of {\sc Turbo} as expected. 
In all scenarios, {\sc DeepTurbo} outperforms both {\sc Turbo} and {\sc NeuralBCJR} on high SNR cases (SNR $\geq$ 0.5 dB), which implies lowered error floor. 
To achieve this performance, it is critical to use appropriate choice of the posterior information dimension $K$. Empirically we find $K=5$ trains faster and also achieves the best performance among all $K<10$. 

\begin{figure}[!ht] 
\centering
\vspace{-0.5em}
\includegraphics[width=0.49\textwidth]{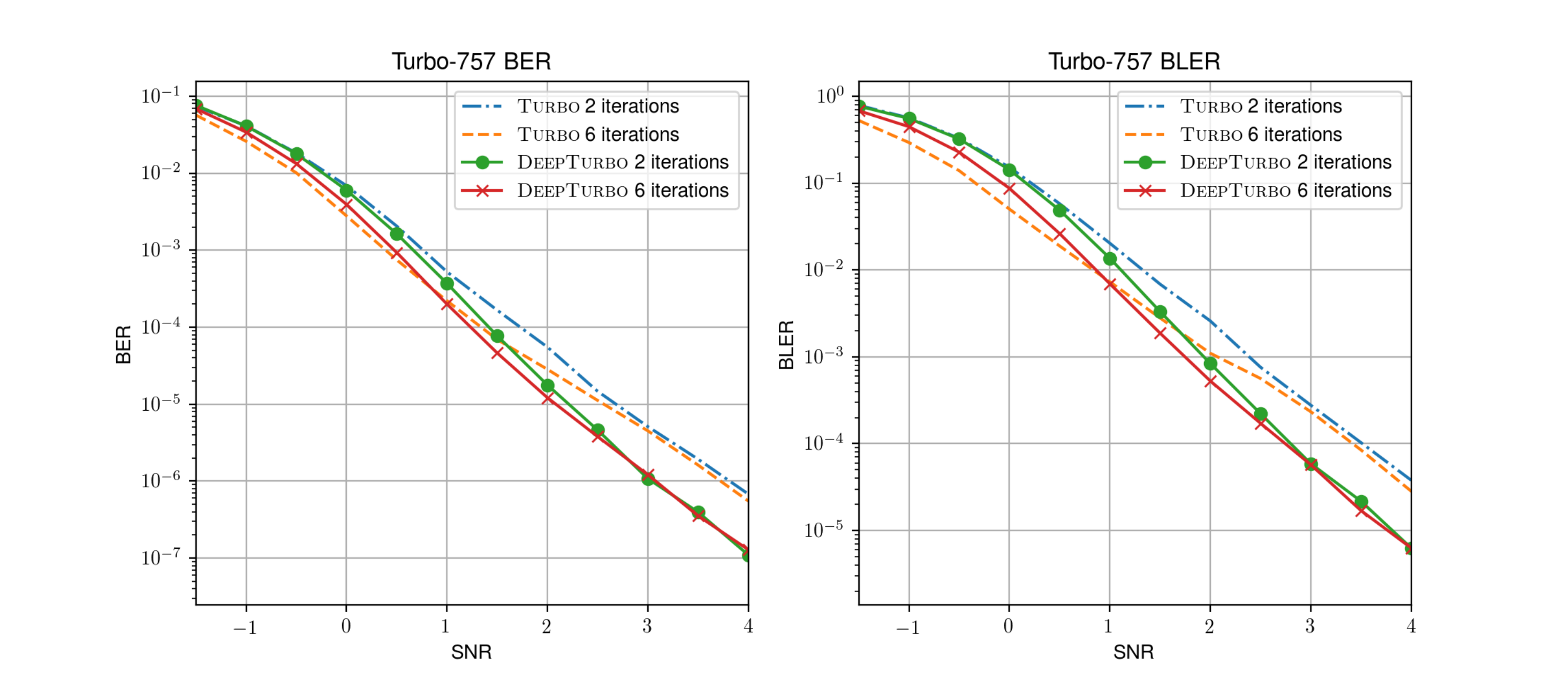}\ \ \ 
\vspace{-1.5em}
\caption{An intermediate layer of {\sc DeepTurbo} already achieves improved performance}
\label{diff_iter}
\vspace{-1.0em}
\end{figure}

In Figure~\ref{diff_iter}, we compare the decoder performances of {\sc DeepTurbo} and {\sc Turbo} with $2$ and $6$ decoding iterations. Compared to {\sc Turbo} with 2 decoding iterations, {\sc DeepTurbo} ($i=2$) shows significant improvement. This implies that the latent representations at lower layers (iterations) of {\sc DeepTurbo} extracts the information faster than iteratively applying BCJR. Hence, {\sc DeepTurbo} can achieve a desired level of accuracy with a smaller number of iterations. 

\subsection{Generalization to Longer Block Lengths}
\begin{figure}[!ht] 
\centering
\vspace{-0.5em}
\includegraphics[width=0.49\textwidth]{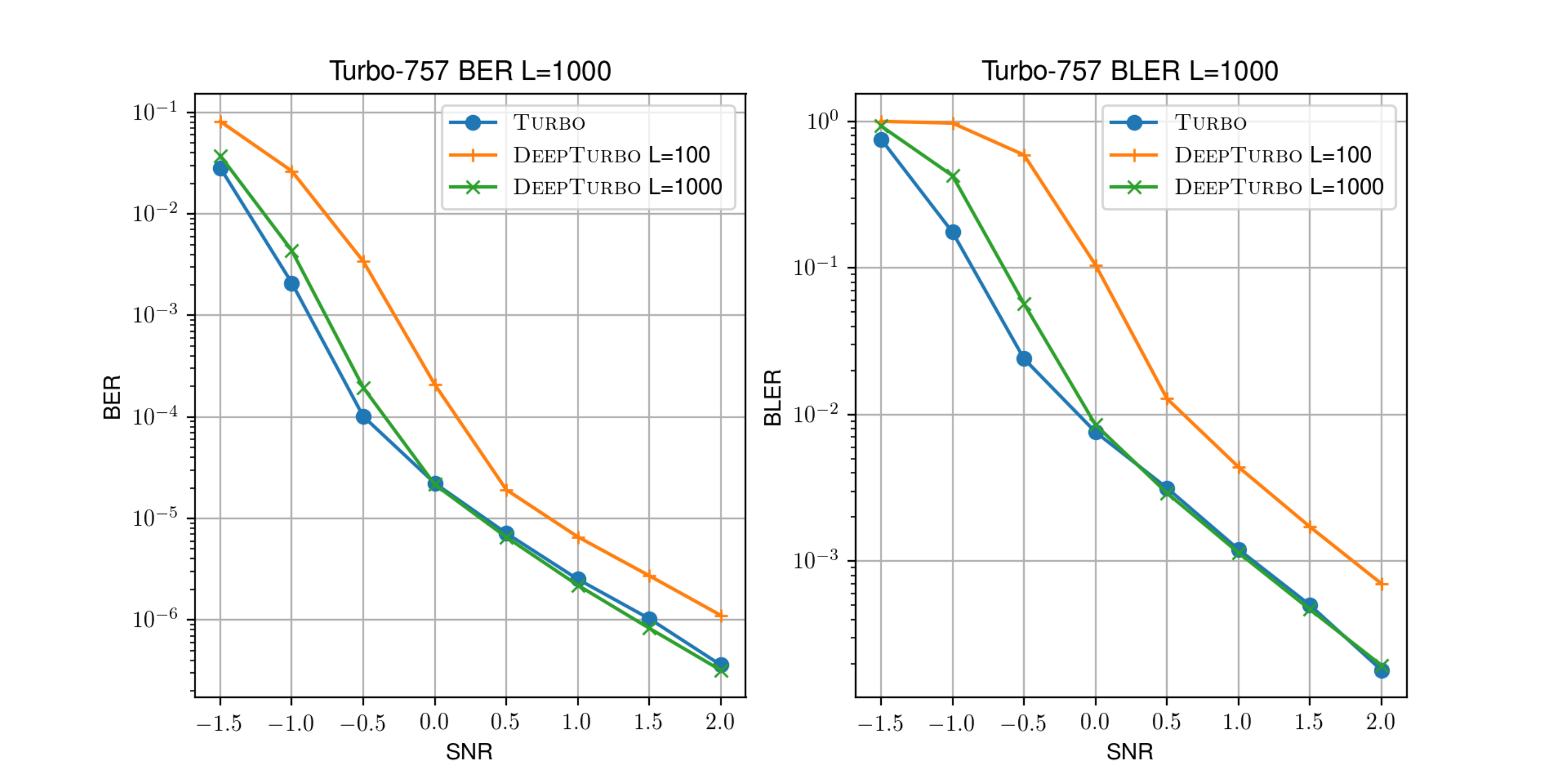}\ \ \ 
\vspace{-1.5em}
\caption{Further training of {\sc DeepTurbo} is required to achieve the desired performance on larger block lengths: BER (left), and BLER (right)}\label{bl1000}
\vspace{-1.0em}
\end{figure}

{\sc Turbo} uses the BCJR algorithm, which is independent of the block length. Hence, {\sc Turbo} is generalizable to  
any block lengths. 
On the other hand, the proposed {\sc DeepTurbo} trained on a short block length ($L=100$) turbo code does not perform as well on a larger block length ($L=1000$) turbo code when applied directly.  This indicates that the gain of {\sc DeepTurbo} in the high SNR regime is due to customizing the decoder to the specific block length it is trained on. We can recover the desired performance on larger block length, by initializing {\sc DeepTurbo} with the model trained on shorter block lengths, and then further training it for small number of epochs. In Figure~\ref{bl1000}, we plot the BER and BLER of the {\sc DeepTurbo} after re-training under block length 1000.

Without relying on the mathematical structure of the code (as exploited in BCJR), generalizing to longer block lengths remains a challenging task. Ideally, we want a decoder that can be trained on short blocks which can be used on longer blocks. 
This will eliminate the bottleneck of several challenges in directly training longer blocks.
For example, training under a large block length requires a large amount of GPU memory. Hence, we cannot train the decoder with large batch sizes, and this results in an unstable training. Furthermore, exploding and diminishing gradient of RNN makes learning unstable. It would be an interesting research direction to design good decoders that generalize to longer block lengths. 

\subsection{Non-AWGN performance}
{\sc DeepTurbo} is tested on the following non-AWGN channels with block length 100:

\begin{itemize}
	\item Additive T-distribution Noise (ATN) channel:  $y = x+z$, where $z \sim T(\nu, \sigma^2)$. 
	\item Radar Channel: $y = x+z+w$. where $z \sim N(0, \sigma_1^2)$ is a background AWGN noise, and $w \sim N(0, \sigma_2^2)$, with probability $p$ is the radar noise with high variance and low probability. $\sigma_1 << \sigma_2$. 
\end{itemize}

\begin{figure}[!ht] 
\centering
\vspace{-0.5em}
\includegraphics[width=0.49\textwidth]{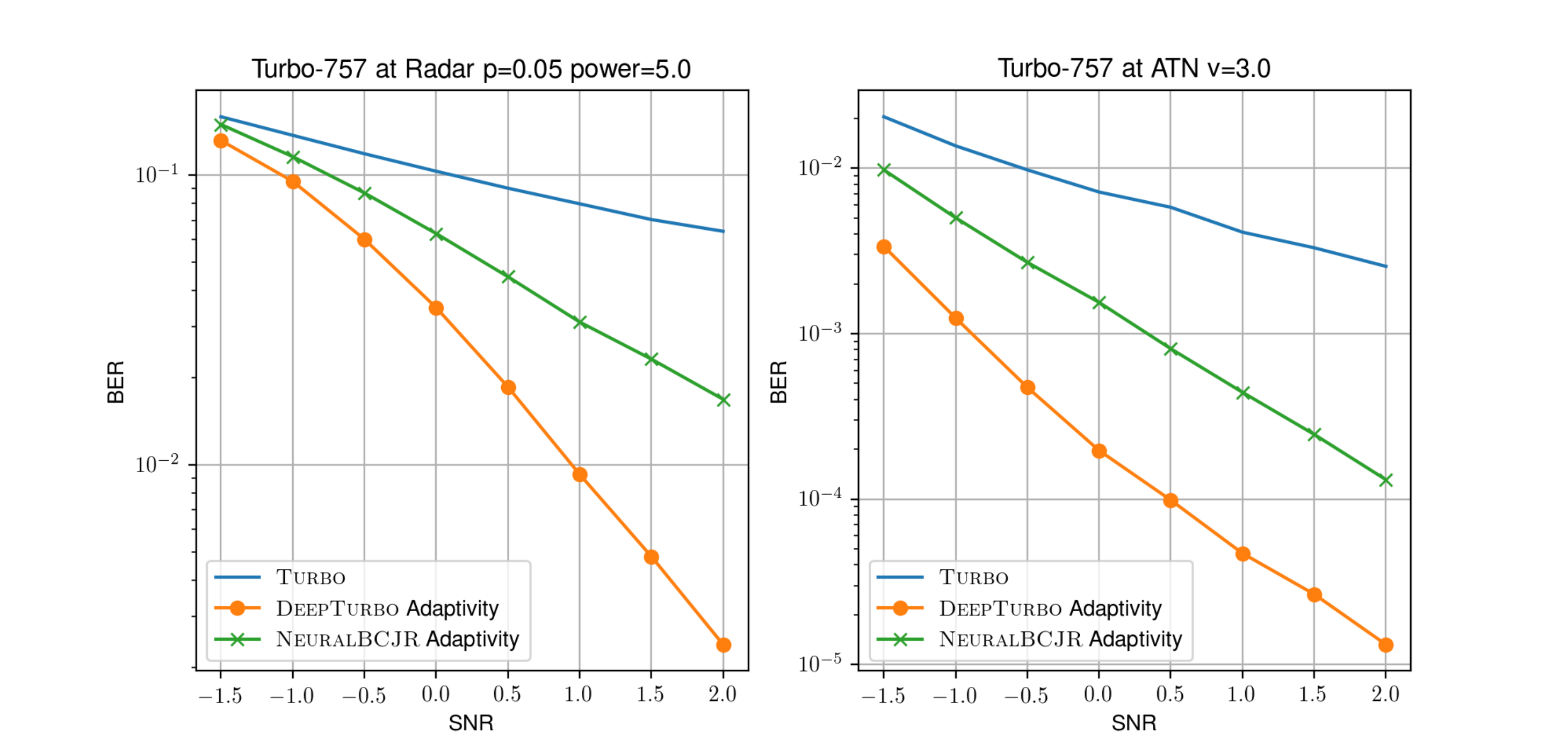}\ \ \ 
\vspace{-1.5em}
\caption{{\sc DeepTurbo} adapts to non Gaussian channels: Radar channel with $p=0.05$ and $\sigma_2 = 5.0$ (left), and ATN with $\nu=3.0$ (right)}\label{better_adapt}
\vspace{-1.0em}
\end{figure}

{\sc NeuralBCJR} shows improved robustness and adaptivity compared to existing heuristics~\cite{kim2018communication}. 
We train {\sc DeepTurbo} on  ATN and Radar end-to-end.
Figure \ref{better_adapt} shows that {\sc DeepTurbo} significantly improves upon {\sc NeuralBCJR}. This is due to the non-shared parameters of {\sc DeepTurbo}, that can perform different decoding functions at different stages of decoding. Hence, {\sc DeepTurbo} has better adaptivity compared to {\sc NeuralBCJR}.

\section{Conclusion}\label{discussion}
We demonstrated that {\sc DeepTurbo}, an end-to-end trained decoder for turbo codes, exhibits an improved reliability, adaptivity and lowered error floor as compared to {\sc NeuralBCJR}, while requiring no knowledge of a BCJR algorithm. 
It has a novel structure, which allows one to achieve required performance with reduced decoding iterations. 
We envision more end-to-end trained neural decoders such as {\sc DeepTurbo} will be proposed in the future for other state-of-the-art codes such as LDPC and Polar codes. We refer the reader to the appendix for further details.

\appendix
\subsection{Deep Learning Modules}\label{appendix_rnns}
\subsubsection{Recurrent Neural network (RNN) and Gated Recurrent Unit (GRU)}
\begin{figure}[!ht] 
\centering
\includegraphics[width=0.40\textwidth]{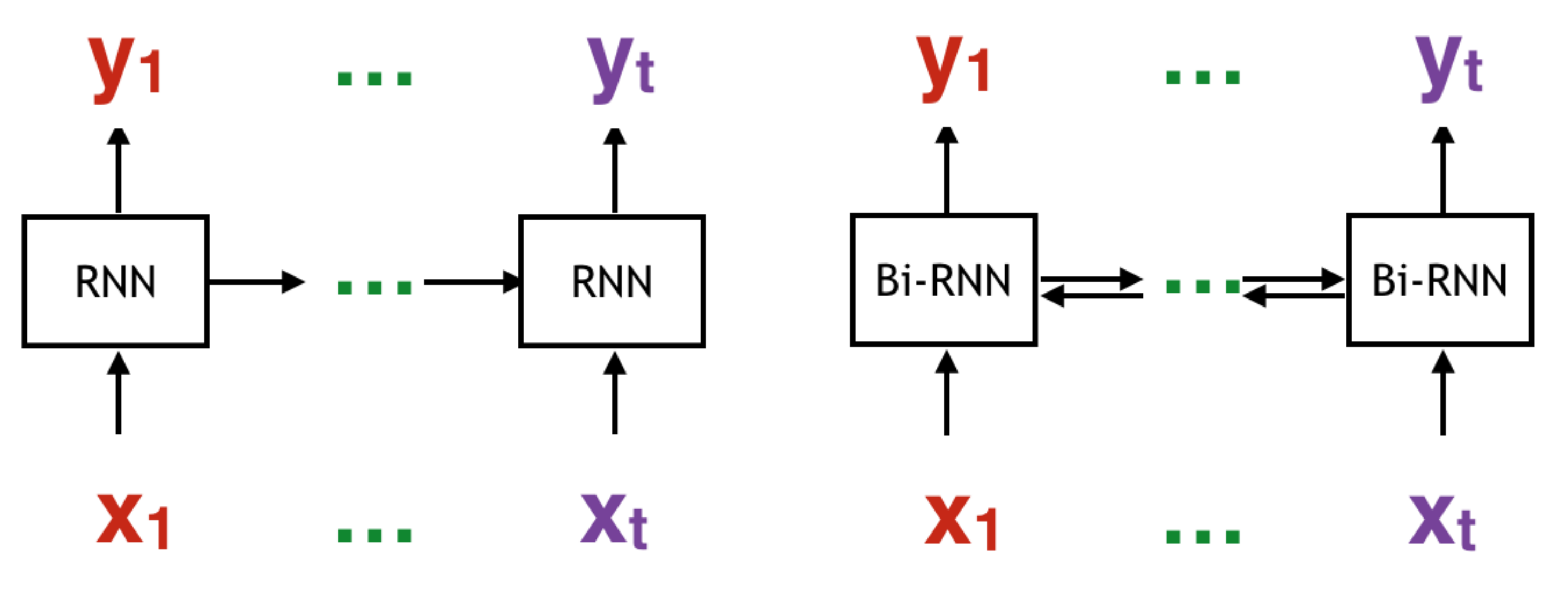}\ \ \ 
\caption{Bi-RNN and RNN}\label{rnns}
\vspace{-0.5em}
\end{figure}

RNN is one of the most widely used deep learning models for sequential data~\cite{goodfellow2016deep}. RNN can be considered as a learnable function $f(.)$ with input $x_t$ and hidden state $h_{t-1}$ at the time $t$, and outputs $y_t$ for next layer and hidden state $h_t$ for the next time. Wide enough RNN theoretically can approximate any function~\cite{schafer2006}~\cite{hornik1989ua}. Bidirectional RNN (Bi-RNN)~\cite{goodfellow2016deep}, combines two RNNs, one for the forward pass and one for the backward pass. can use both information from the past and future. Bi-RNN can be considered as a generalization of forward-backward algorithm such as the BCJR algorithm~\cite{kim2018communication}. RNN and Bi-RNN are shown in Figure \ref{rnns}.

\begin{figure}[!ht] 
\centering
\includegraphics[width=0.40\textwidth]{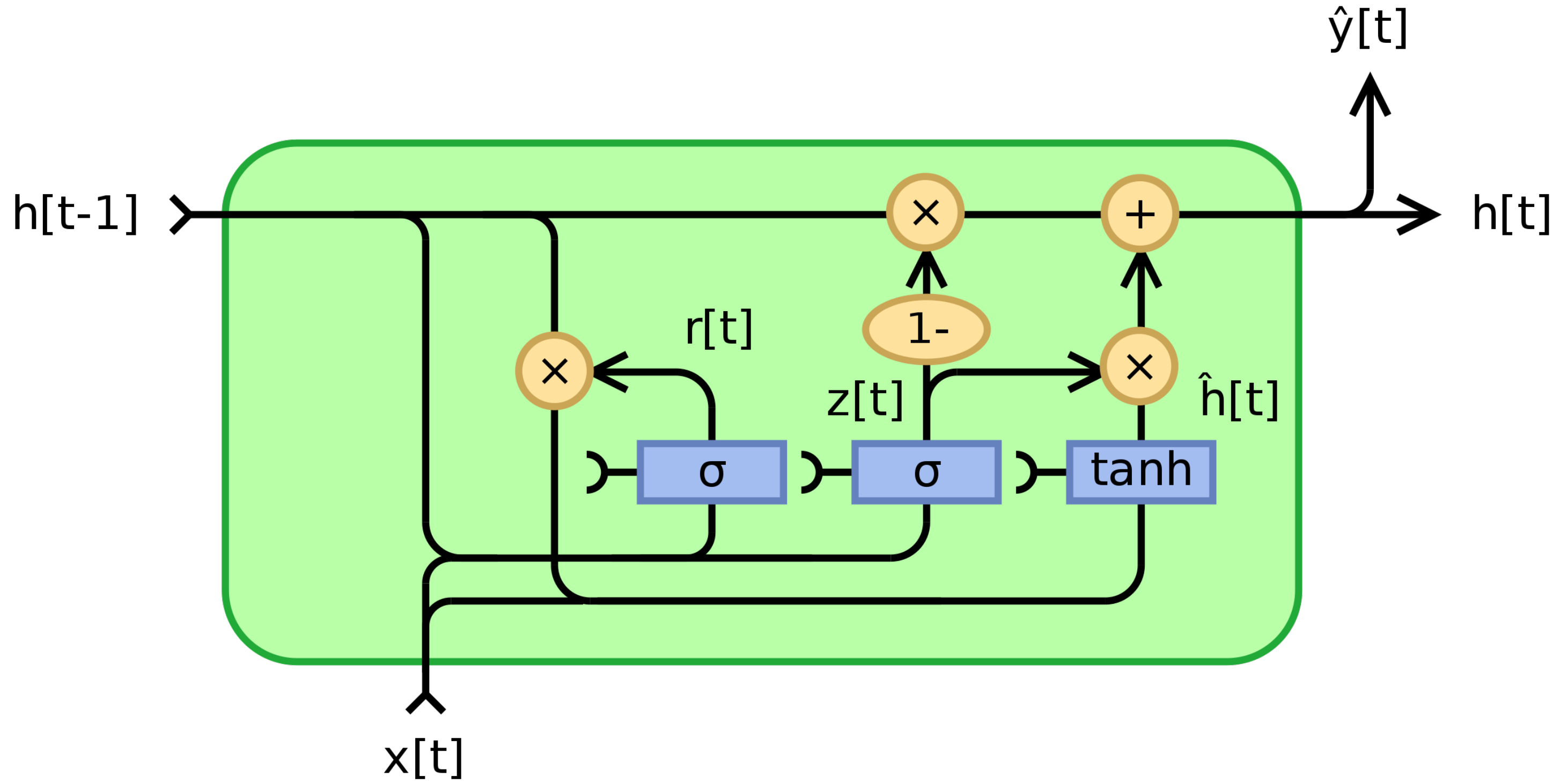}\ \ \ 
\caption{GRU}\label{gru}
\vspace{-0.5em}
\end{figure}

RNN are not compatible to capture long-term dependencies due to exploding gradient problem, in practice Long Short Term Memory (LSTM) and Gated Recurrent Unit (GRU)~\cite{goodfellow2016deep} are used to alleviate gradient exploding problem via gating schemes. GRU~\cite{chung2014empirical} is shown in Figure \ref{gru}. This paper use Bidirectional GRU (Bi-GRU) as the primary learnable structure. 

\subsubsection{Same Shape 1D-CNN}\label{cnns}
1D Convolutional Neural Network (1D-CNN) is widely used for sequential data with limited dependency length~\cite{goodfellow2016deep}. 1D-CNN takes the input of shape $(L, F1)$, with data block length $L$, and output feature size $F2$, and use $F2$ filters to conduct convolution with kernel size $k$. After convolving with kernel of odd size $k$, with each time the convolution kernel moves ahead by 1 step (termed as stride = 1) and the output length becomes $L  - k + 1$. 

\begin{figure}[!ht] 
\centering
\includegraphics[width=0.40\textwidth]{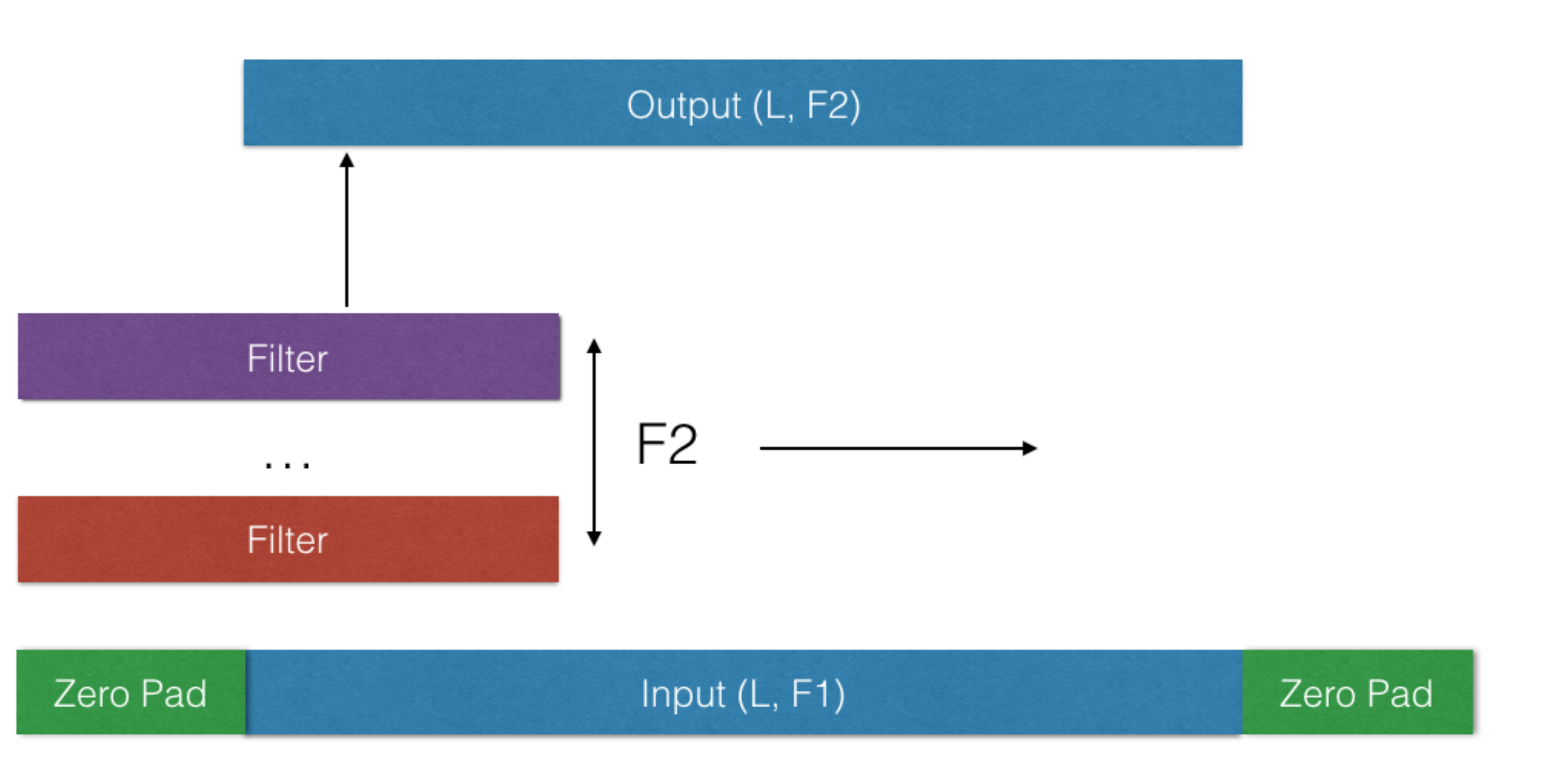}\ \ \ 
\caption{Same Shape 1D CNN with 1 layer}\label{sameshapecnn}
\vspace{-0.5em}
\end{figure}

Same shape 1D-CNN means that the input sequence and output sequence have the same length $L$.  Same shape 1D-CNN has stride = 1, for each odd kernel size $k$, the zero padding is $\frac{k-1}{2}$ for both the beginning part and the ending part, such that the input and output block length $L$ is preserved~\cite{goodfellow2016deep}, as shown in Figure \ref{sameshapecnn}.

In this paper, we stack multiple same shape 1D-CNNs to construct a deep CNN based SISO decoder. Same shape 1D-CNN doesn't have a long dependency, we expect better trainability and slightly degraded performance. We use 5 layers, 100 filters same shape 1D-CNN, with kernel size 5. Empirically, the memory usage of 5 layer 100 filter same shape 1D-CNN is about 1/10 comparing to 2 layer 100 unit Bi-GRU, while the training time is about 1/5 of 2 layer 100 units Bi-GRU. 

\subsection{{\sc DeepTurbo-CNN}}
\label{cnn}

RNNs are hard to parallelize for long blocks due to their sequential dependency structure~\cite{hwang2015rnnpara}. RNN is also hard to train due to exploding gradient problem~\cite{chung2014empirical}. Replacing RNN with Convolutional Neural Network (CNN) has been proposed in recent NLP research~\cite{seo2016biaf}. Inspired by the success of CNNs for sequential data, we propose {\sc DeepTurbo-CNN}, which simply replaces the Bi-GRU of {\sc DeepTurbo} by multi-layer same shape 1D-CNN discussed in previous section. Since CNN has limited and fixed dependency length, we expect {\sc DeepTurbo-CNN} to have  degraded performance comparing to {\sc DeepTurbo}. However, since CNN can be efficiently computed and trained with much less computation and memory, {\sc DeepTurbo-CNN} shows promise for future deployment.

\begin{figure}[!ht] 
\centering
\vspace{-1.0em}
\includegraphics[width=0.49\textwidth]{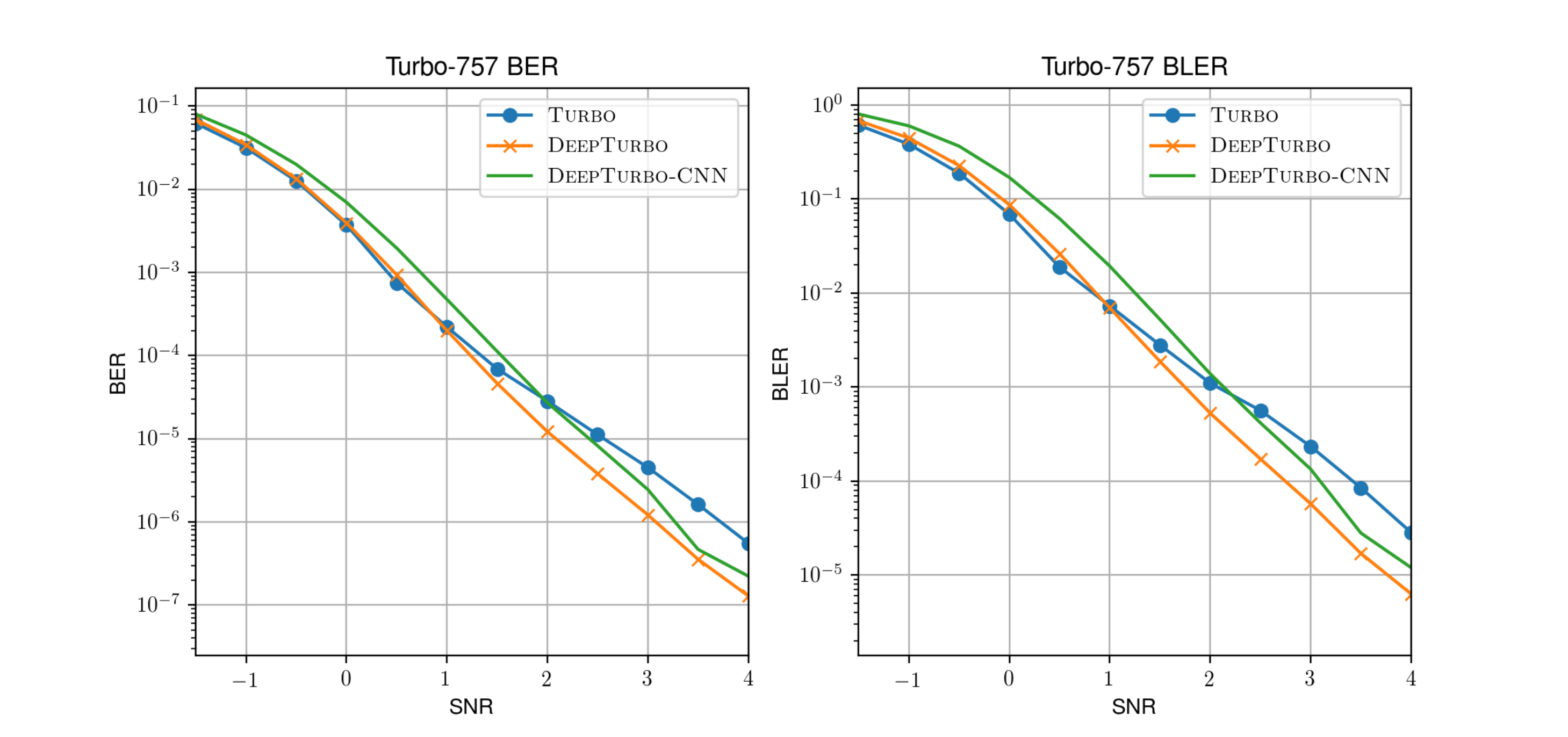}\ \ \ 
\vspace{-1.5em}
\caption{{\sc DeepTurbo-CNN} performance slightly degrades at the gain of computational efficiency on AWGN channel: BER (left) and BLER (right)}\label{cnn_turbofy}
\vspace{-1.0em}
\end{figure}

We test the performance of {\sc DeepTurbo-CNN} with the same setup as Section~\ref{perf}, with block length $L=100$ and 6 decoding iterations. 
In Figure \ref{cnn_turbofy}, {\sc DeepTurbo-CNN} exhibits a 0.5dB degradation compared to {\sc DeepTurbo}. 
{\sc DeepTurbo-CNN} requires much less computation and memory and is more efficient to run in parallel computing environments than {\sc DeepTurbo}. 

\subsection{{\sc DeepTurbo} Network Design and Training Methodology}\label{appendix_hyper}
In this section, we discuss about the hyper-parameter design and training methodology of {\sc DeepTurbo}. We use either BER or learning curve to compare different designs of {\sc DeepTurbo}. The learning curve shows the validation loss dynamics during training, the x-axis is the training time measured in epoch, and the y-axis is the validation loss. 
The desired features are lower validation loss and faster convergence.

\subsubsection{Optimal Information Feature Size $K$}
\begin{figure}[!ht] 
\centering
\includegraphics[width=0.40\textwidth]{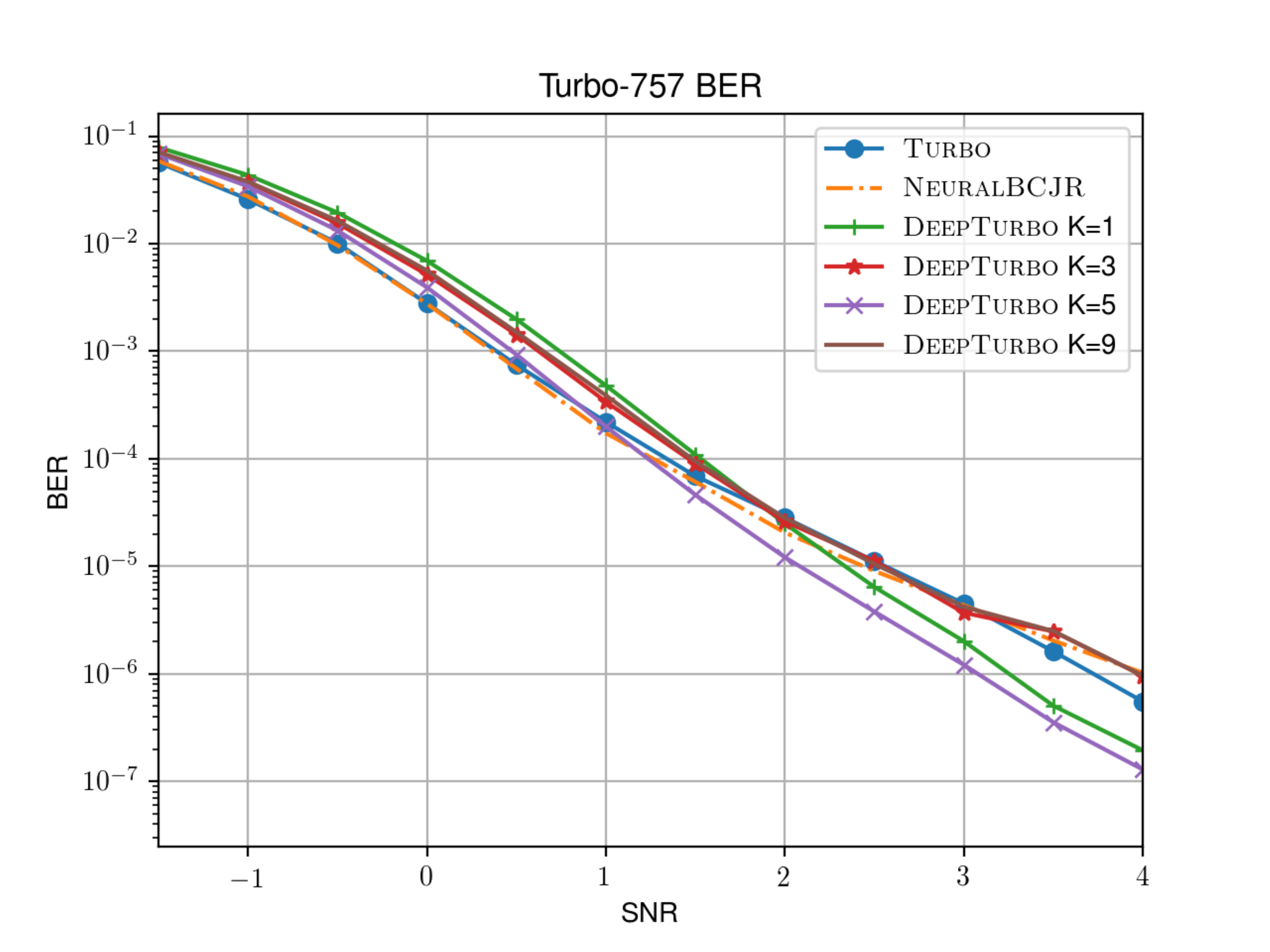}\ \ \ 
\caption{$K=5$ shows best performance among different selections}\label{keffect}
\vspace{-0.5em}
\end{figure}

{\sc DeepTurbo} performance is affected by information size $K$. Testing on different $K$ is shown in Figure \ref{keffect}. Among different $K$, $K=5$ shows the best performance. Smaller $K$ limits the information transferred to the next stage, while larger $K$ is harder to train.

\subsubsection{Residual Connection improves training {\sc DeepTurbo}}
\begin{figure}[!ht] 
\centering
\includegraphics[width=0.40\textwidth]{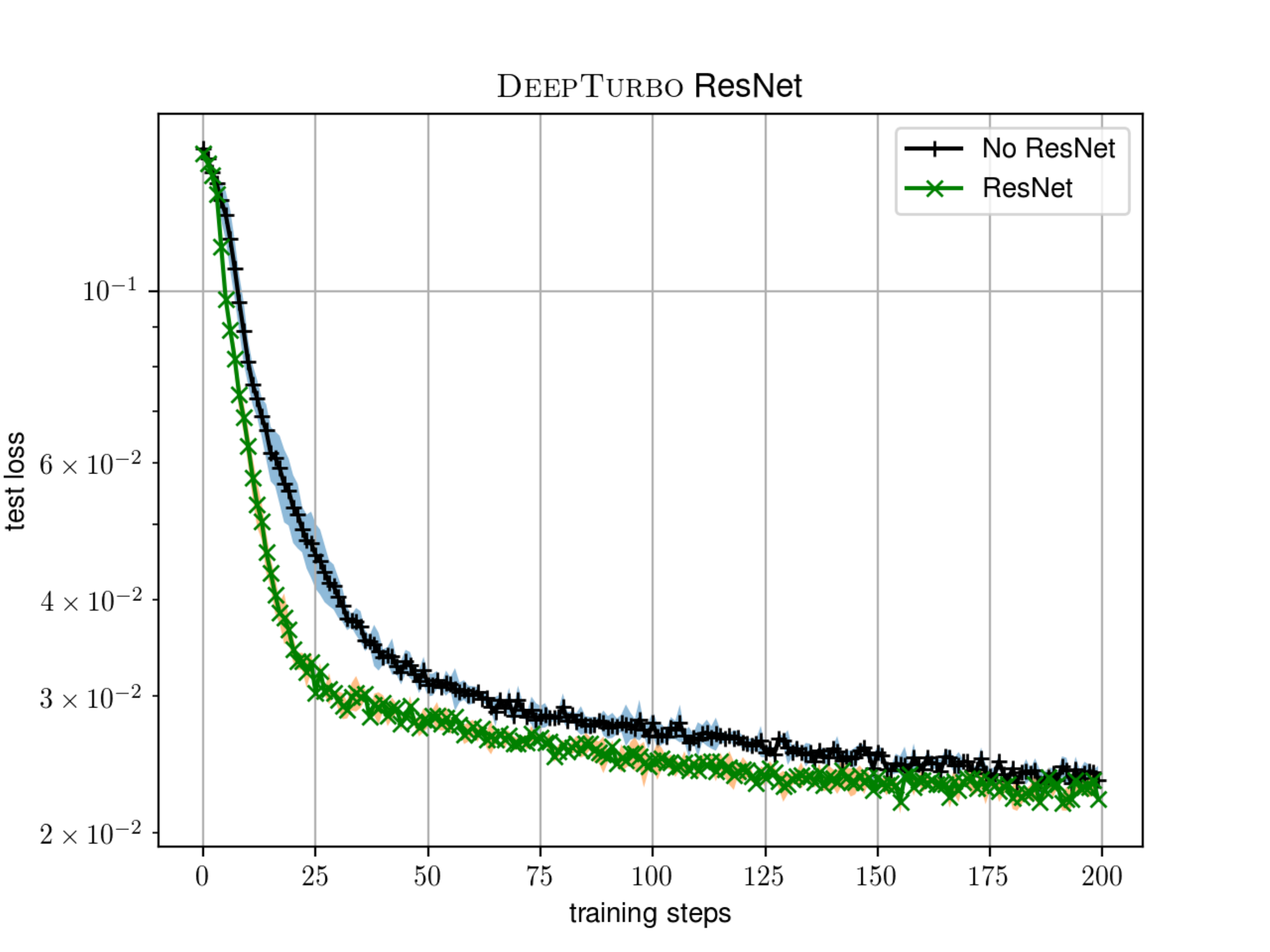}\ \ \ 
\caption{Improved training with ResNet }\label{resnet}
\vspace{-0.5em}
\end{figure}

Residual Neural Network (ResNet)~\cite{he2016resnet} uses short-cut to improve the training of deep networks. Training {\sc DeepTurbo} with residual connection not only imitates extrinsic information via existing Turbo decoder, but also improves training and shows a better performance, as shown in Figure \ref{resnet}.

\subsubsection{Larger network learns faster}
\begin{figure}[!ht] 
\centering
\includegraphics[width=0.40\textwidth]{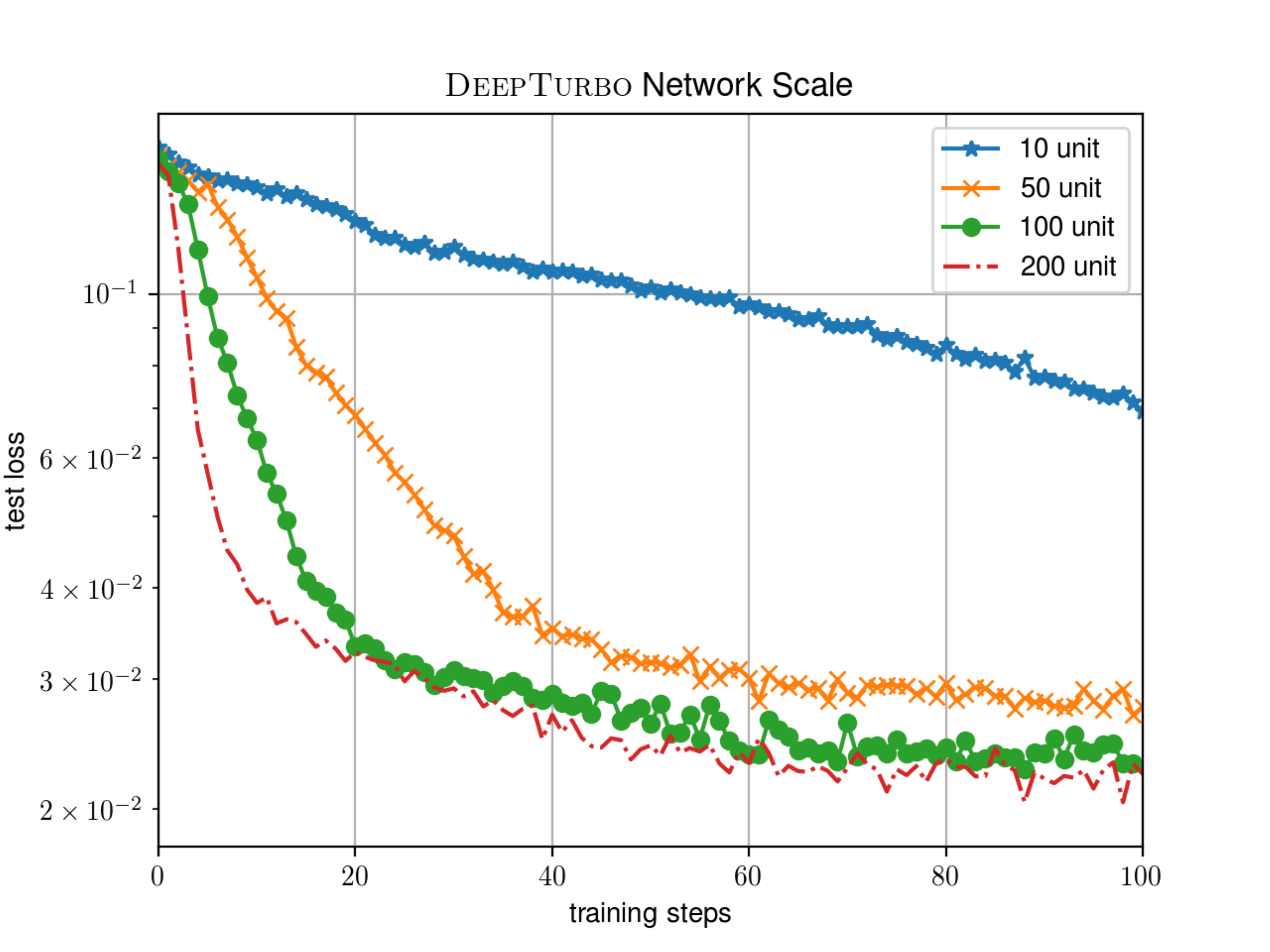}\ \ \ 
\caption{Learning curve of block length $L=20$, larger network learns faster}\label{largenet}
\vspace{-0.5em}
\end{figure}

Larger network with more units makes training faster, as shown in Figure \ref{largenet}.  We choose 100 units due to computation limit. 100 units already show a better performance compared to {\sc NeuralBCJR}, we expect even better performance with larger {\sc DeepTurbo} models depending upon the availability of computation resources.

\subsubsection{{\sc DeepTurbo} has better training stability than {\sc NeuralBCJR}}
\begin{figure}[!ht] 
\centering
\includegraphics[width=0.40\textwidth]{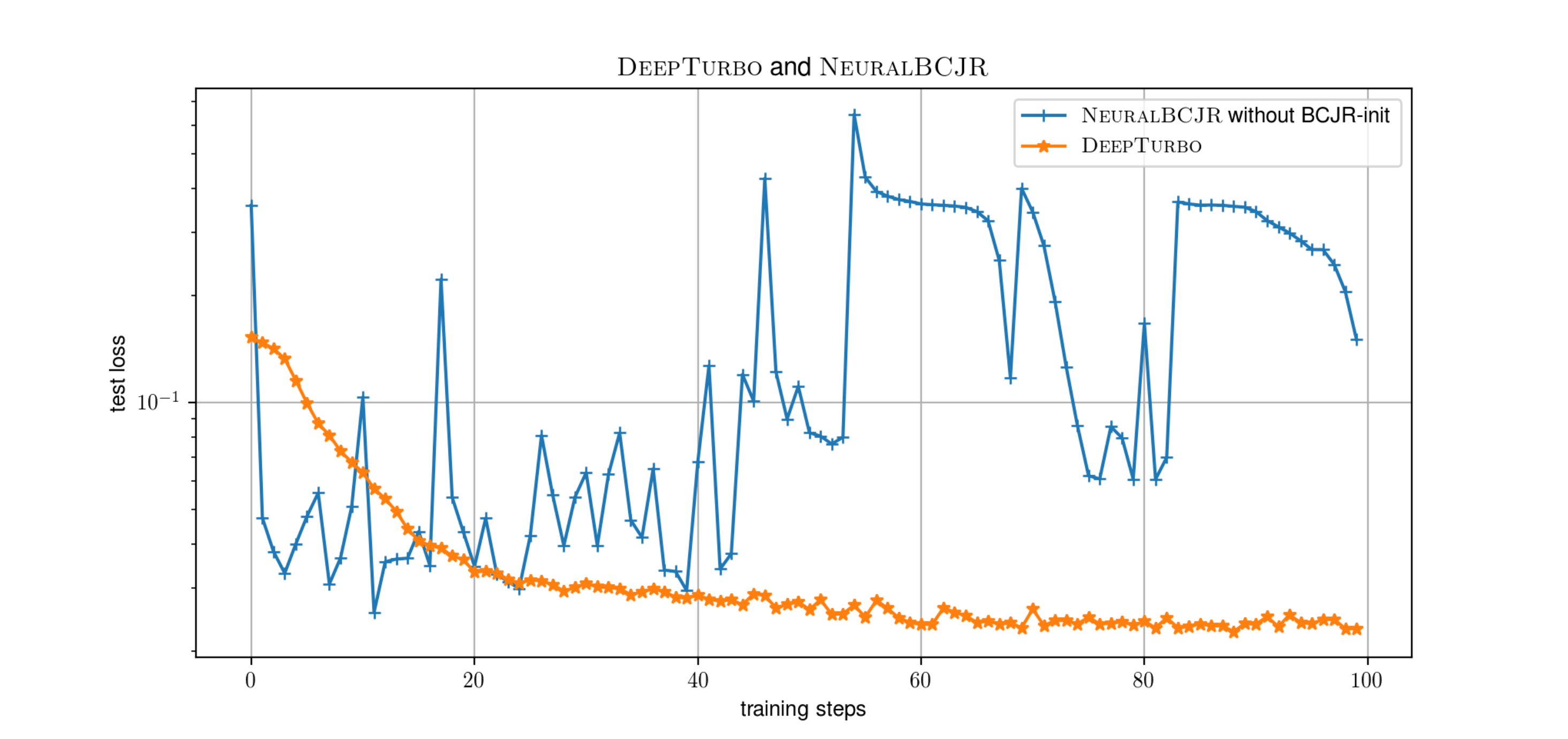}\ \ \ 
\caption{{\sc DeepTurbo} shows better trainability than {\sc NeuralBCJR} without BCJR initialization}\label{ntd}
\vspace{-0.5em}
\end{figure}

{\sc NeuralBCJR} without BCJR initialization is very unstable during training, while {\sc DeepTurbo} shows progressive improvement as shown in Figure \ref{ntd}. Both {\sc NeuralBCJR} and {\sc DeepTurbo} use Adam optimizer with learning rate 0.001.

\subsubsection{Training SNR}
\begin{figure}[!ht] 
\centering
\includegraphics[width=0.40\textwidth]{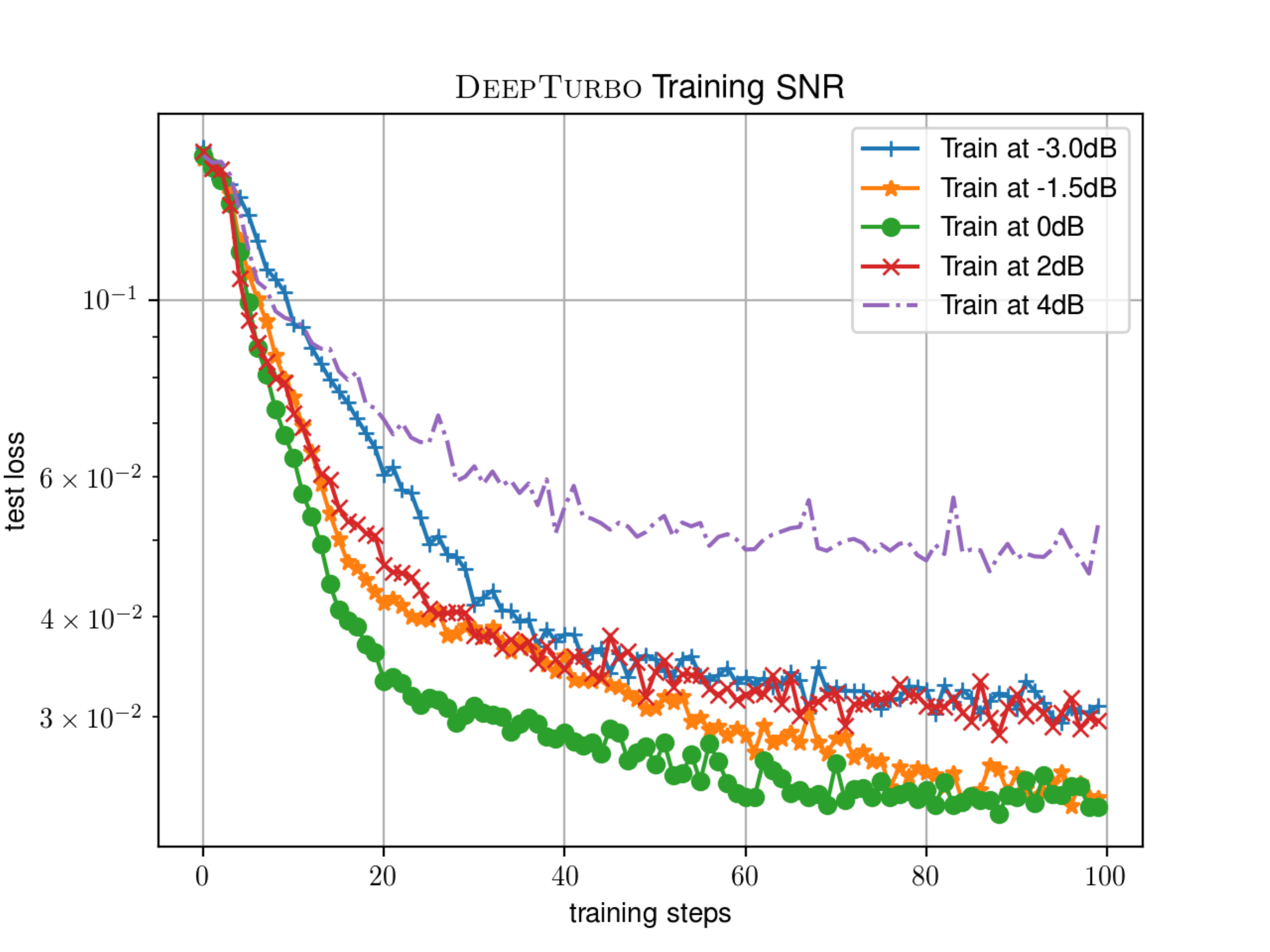}\ \ \ 
\caption{Optimal Training SNR}\label{trainsnr}
\vspace{-0.5em}
\end{figure}

Minimizing the Bit Error Rate (BER) of the neural decoder can be considered as a sequence of classification problems, where typically Binary Cross-Entropy (BCE) loss is used as a differentiable surrogate loss function~\cite{goodfellow2016deep}. Training decoder under low SNR close to Shannon Limit results in best empirical performance~\cite{kim2018communication}, since adding Gaussian noise is equivalent to adding regularizer to the neural decoder~\cite{jiang2018learn}. Here we examine the effect of different training levels of SNR on  {\sc DeepTurbo}. The learning curve is shown in Figure \ref{trainsnr}. 

Training with 0dB shows fastest convergence, while training with -1.5dB shows slightly better validation loss when converge. So we schedule training SNR by starting to train at 0dB, and fine-tune at -1.5dB after convergence. 

\subsubsection{{\sc DeepTurbo} Reduces Decoding Iterations}

Interpretation of the $K$ dimensional latent representation is harder than interpreting log-likelihood (LLR) with dimension one. We examine the evolution of posteriors of each stage by auxiliary classifier method~\cite{duan2009auxclass}. After {\sc DeepTurbo} converges, we fix the weights of all of its layers. For a given iteration $i \in \{1,...,6\}$ with output posterior of shape $(L,K)$, we add an auxiliary linear layer of shape $(K,1)$ followed by sigmoid activation function to decode. We train auxiliary linear layer with posterior of shape $(L,K)$ as input, and message $(L,1)$ as output till convergence. Training the auxiliary linear layer harvests the information of refined posterior of dimension $K$. Figure \ref{diff_iter} uses the auxiliary classifier method.

\subsection{Open issues and future works}
\subsubsection{{\sc DeepTurbo} Performance at Low SNR}
{\sc DeepTurbo}'s performance at low SNR is not as good as {\sc NeuralBCJR} and the canonical Turbo decoder. The high SNR coding gain of {\sc DeepTurbo} is better, while at low SNR the coding gain is smaller. Designing neural decoder to work well for all levels of SNR is a direction for future research.

\subsubsection{{\sc DeepTurbo} Optimizing BLER Performance}
When the objective is to minimize Block Error Rate (BLER), training with BCE loss doesn't directly minimize the BLER. Using max-BCE loss minimizes the maximum BCE loss along with the whole sequence of length $L$: $max-BCE = \max_{i \in \{1,...,L\}} BCE(f(x_i), y_i)$. As the max-BCE has sparser gradients, our empirical results show minimal gain for BLER. Improving BLER performance is an interesting future direction.

\subsubsection{Design Better Learnable Structure}
Although both turbo-LTE and turbo-757 performances are shown to have desired features, hyper-parameter tuning and training on turbo-LTE are experiencing much harder than that on turbo-757. Also training with longer block length is very time consuming. RNN has the inevitable gradient exploding problem, while CNN has limited and fixed dependency length. Designing a better learnable structure which improves both the capacity and the learnability is an interesting future research direction, which will have ramifications beyond channel coding community.

\subsubsection{Model Reduction}
{\sc DeepTurbo} has multiple iterations. Although non-shared weights across different decoding stages result in better performance, it is still favorable for the sake of reduced storage and deployment complexity to reduce the number of weights. One possible direction is to make part of model between each iteration share their parameters, thus can significantly reduce the number of weight to be saved for decoding model.

\subsubsection{{\sc DeepTurbo} Inspired Decoding Algorithm}
As {\sc DeepTurbo} learns a neural decoder with better performance, one potential research direction is to distill what {\sc DeepTurbo} has learnt, and implement the algorithm to improve existing Turbo decoders.

\subsubsection{Compare to State-of-the-art Error Floor Lowering Techniques}
{\sc DeepTurbo} works well under short block lengths. However most error floor lowering results are on longer block length ($L \geq 528$). Comparing these results with existing state-of-the-art error floor lowering scheme can be an interesting and impactful future research direction.


 
\end{document}